\documentclass[a4paper,11pt]{article}

\usepackage{pos}
\usepackage{hyperref,subcaption}
\usepackage{lipsum} 

\usepackage{setspace}

\title{Multimessenger Potential of the Radio Neutrino Observatory in Greenland}
\author*[a,b,c]{Marco Stein Muzio} 
\onbehalf{{for the RNO-G Collaboration}\\{\normalsize \normalfont(a complete list of authors can be found at the end of the proceedings)}\\}

\affiliation[a]{Department of Physics, Pennsylvania State University, University Park, PA 16802, USA}
\affiliation[b]{Department of Astronomy and Astrophysics, Pennsylvania State University, University Park, PA 16802, USA}
\affiliation[c]{Institute of Gravitation and the Cosmos, Center for Multi-Messenger Astrophysics, Pennsylvania State University, University Park,
PA 16802, USA}
\emailAdd{msm6428@psu.edu}

\ShortTitle{Multimessenger Potential of RNO-G}

\abstract{
The Radio Neutrino Observatory in Greenland (RNO-G) is the only ultrahigh energy (UHE, ${\gtrsim}30$~PeV) neutrino monitor of the Northern sky and will soon be the world's most sensitive high-uptime detector of UHE neutrinos. Because of this, RNO-G represents an important piece of the multimessenger landscape over the next decade. In this talk, we will highlight RNO-G's multimessenger capabilities and its potential to provide key information in the search for the most extreme astrophysical accelerators. In particular, we will highlight opportunities enabled by RNO-G's unique field-of-view, its potential to constrain the sources of UHE cosmic rays, and its complementarity with IceCube at lower energies.
}

\ConferenceLogo{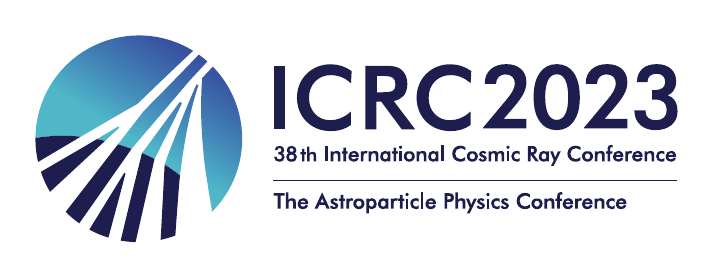}

\FullConference{The 38th International Cosmic Ray Conference (ICRC2023)\\ 26 July -- 3 August, 2023\\ Nagoya, Japan}

\begin{document}

\maketitle

\section{Introduction}\label{sec:intro}

\par
The Radio Neutrino Observatory in Greenland (RNO-G,~\cite{RNO-G:2020rmc}) is an in-ice radio experiment located at Summit Station, Greenland. RNO-G is designed to detect ultrahigh energy (UHE, ${\gtrsim} 30$~PeV) neutrinos. UHE neutrinos are predicted to be produced by UHE cosmic rays (UHECRs), whose origins are still unknown. In particular, photopion production interactions of UHECRs with the cosmic microwave background (CMB) -- often referred to as the Greisen-Zatsepin-Kuzmin (GZK) effect~\cite{Greisen:1966jv,Zatsepin:1966jv} --- imprints a horizon of ${\sim} 100$~Mpc on UHECRs above ${\sim}10^{19.7}$~eV. This makes it impossible to study UHECRs beyond the GZK horizon directly. However, UHE neutrinos (produced in the decay of such photopions) propagate through the universe unimpeded, suffering only redshift losses. This makes UHE neutrinos both a smoking gun of UHECR production and a window into the extreme astrophysical universe on cosmological scales. 

\begin{figure}
    \centering
    \includegraphics[height=3.75in]{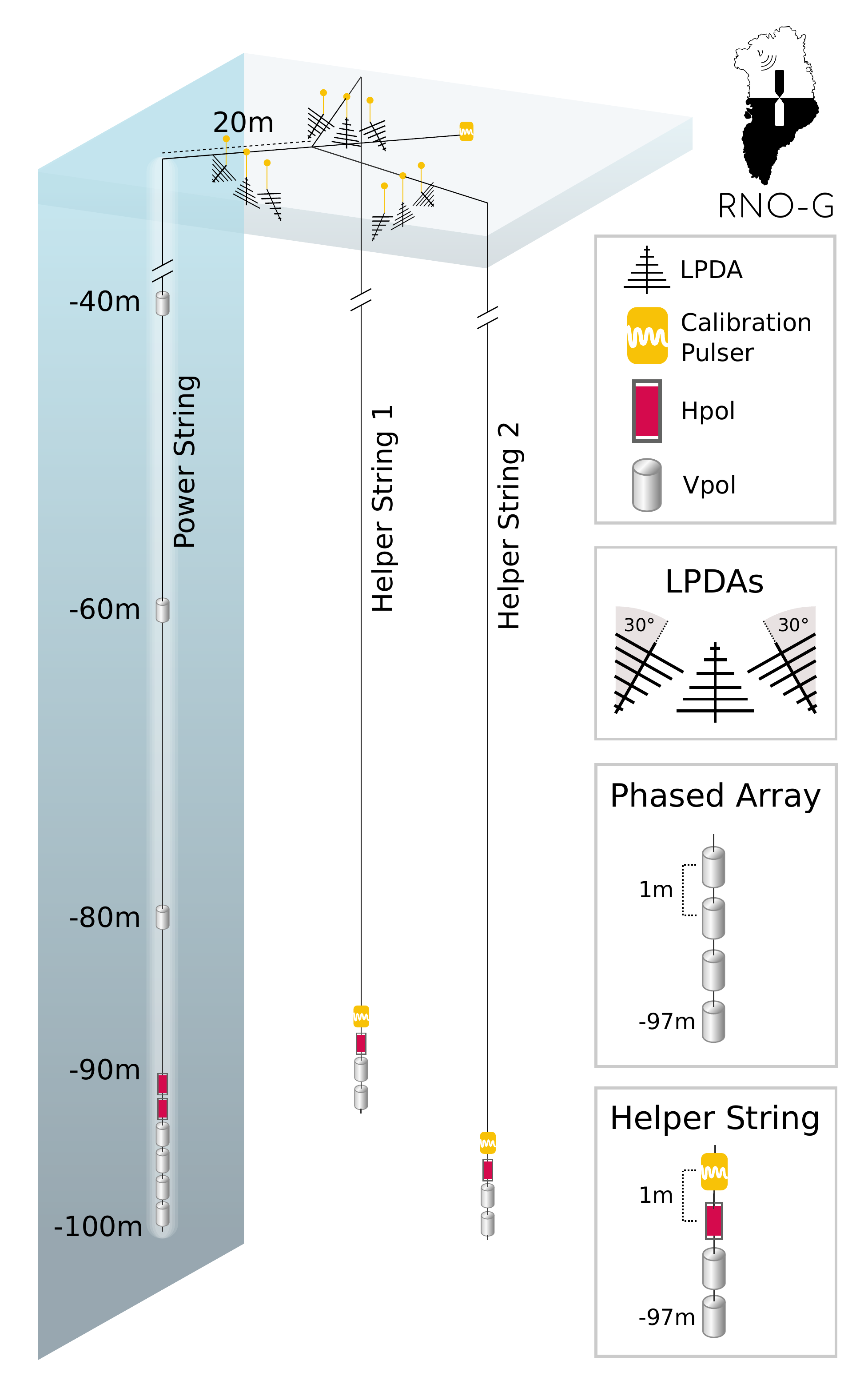}
    \includegraphics[height=3.25in]{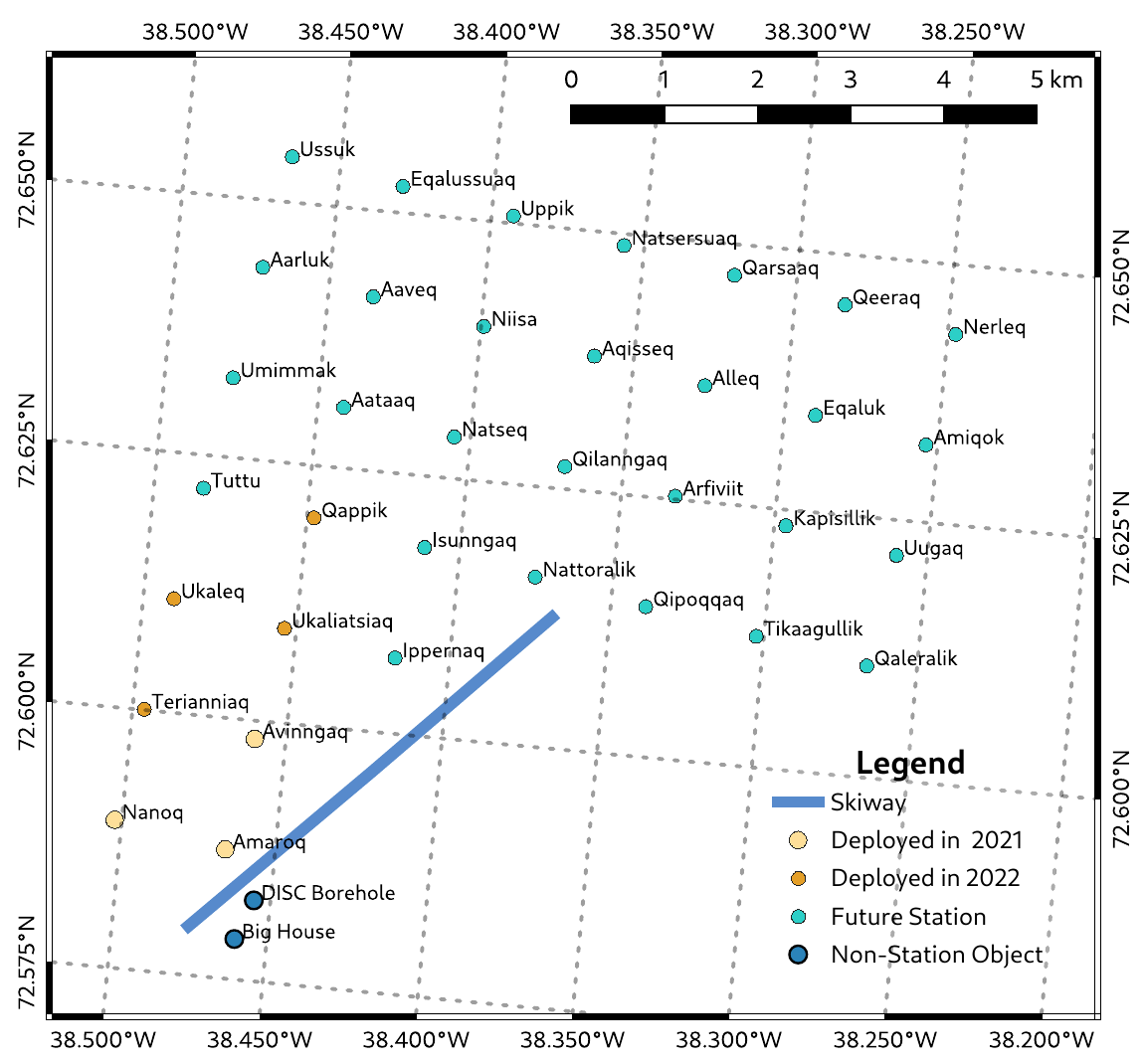}
    \caption{Left: Layout for a single RNO-G station. Right: Map of RNO-G's 35 station array.}
    \label{fig:station}
\end{figure}

\par
If a UHE neutrino has an interaction as it traverses the Earth, it will initiate a particle shower. Especially if this shower develops in a dense medium, like glacial ice, a charge asymmetry will develop at the shower front leading to the emission of radio waves. Radio waves have an exceptionally long attenuation length in ice, on the order of ${\sim} 1$~km, allowing for radio receivers embedded in ice to efficiently monitor a large volume. RNO-G takes advantage of this detection principle. 

\par
Currently, RNO-G has 7 stations deployed and taking data, and upon completion in 2027 will consist of 35 independent stations. Stations of RNO-G are separated by ${\sim} 1.25$~km with the entire array encompassing $40$~km\textsuperscript{2}. The array and station layout are illustrated in Fig.~\ref{fig:station}. RNO-G stations employ a hybrid design, taking advantage of both surface and deep antennas. Deep antennas are distributed across three strings embedded into the ice: one ``power'' string and two ``helper'' strings. The power string consists of 9 antennas (7 sensitive to vertically polarized signals (Vpols) and 2 to horizontally polarized signals (Hpols)): 3 spread between $40$~m and $80$~m in depth and 6 closely spaced at ${\sim}100$~m depth. The two helper strings each consist of 3 closely space antennas (2 Vpols and 1 Hpol) at ${\sim} 100$~m depth. The deep antennas provide improved sensitivity, with the power string providing a phased trigger and the helper strings providing additional reconstruction power. Each station also has 9 surface antennas with an independent trigger, which provide improved event reconstruction and background rejection. This hybrid design will allow RNO-G to lead the next generation of UHE neutrino observatories, balancing precision pointing for multimessenger follow-up with an unprecedented diffuse flux sensitivity.

\section{Role in multimessenger landscape over the coming decade}\label{sec:MMlandscape}

\par
RNO-G's location at in the Northern hemisphere makes it unique for both in-ice radio neutrino observatories and, in particular, for ultrahigh energy neutrino observatories. While other neutrino observatories exist in the Northern hemisphere, including ANTARES~\cite{ANTARES:2011hfw}, Baikal-GVD~\cite{Baikal-GVD:2021zsq}, KM3Net~\cite{KM3Net:2016zxf}, and P-ONE~\cite{P-ONE:2020ljt}, none of these are particularly sensitive to neutrinos above $30$~PeV. Similarly, despite being located at the South Pole, IceCube has a view of the Northern sky but only at lower energies where the Earth is transparent to neutrinos. 

\begin{figure}
    \centering
    \includegraphics[width=0.9\textwidth]{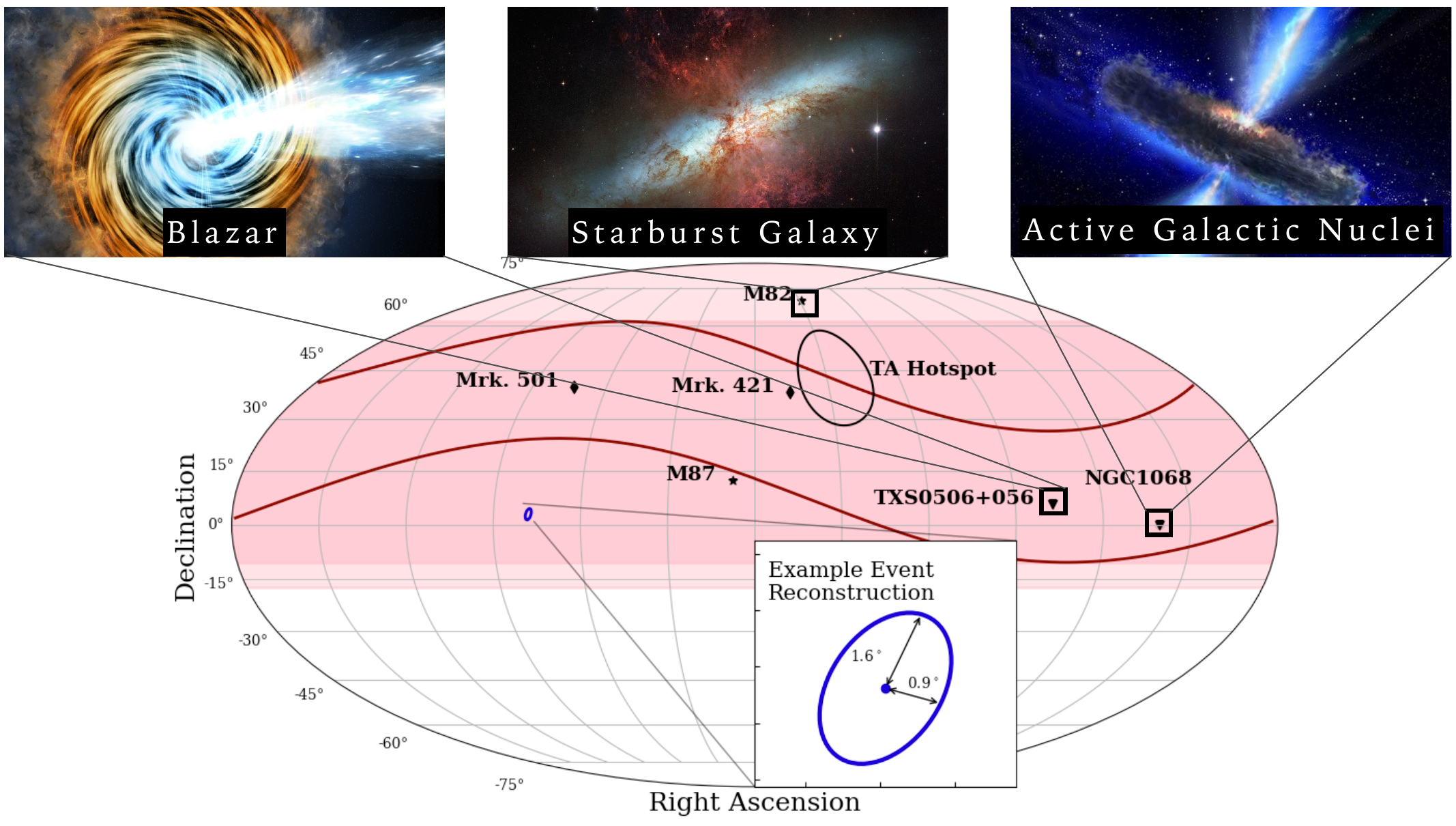}
    \caption{Expected field-of-view of RNO-G along with a number of notable point sources. Red lines indicate RNO-G's instantaneous sky coverage while bands indicate the daily sky coverage in a narrow (dark) and wide (light) altitude range. An example high-quality event reconstruction is also shown.}
    \label{fig:skymap}
    \vspace{-0.1in}
\end{figure}

\par
The Northern sky is a particularly interesting region to explore, both because at ultrahigh energies it is the least explored region of the neutrino sky and because it contains a number of interesting possible neutrino sources (as shown in Fig.~\ref{fig:skymap}). First and foremost, the only known extragalactic sources of neutrinos, TXS-0506+056 and NGC 1068, are located in the Northern sky~\cite{IceCube2018b, IceCube:2018dnn,IceCube:2022der}. NGC 1068 is a particularly interesting source since it is the first known point source of neutrinos. Further, to what degree the hard spectrum of TXS-0506+056 continues to higher energies, making it an emitter of UHE neutrinos, is an open question. RNO-G will be the only observatory currently planned which will be able to address these questions.

\par
While no other sources of neutrinos are currently known, the Northern sky contains a number of other extreme astrophysical sources which are good candidates to be neutrino sources. These include bright starburst galaxies, like M82, and nearby blazars, like Mrk421 and Mrk501. Additionally, there is some evidence for significant UHE cosmic ray (UHECR) production in the Northern sky. The Telescope Array Collaboration has reported two excesses (hotspots) of UHECRs from the Northern sky~\cite{TAhotspots}. Additionally, models of the UHECR dipole observed by the Pierre Auger Observatory~\cite{AugerDipole} also indicate that the Virgo cluster (of which M87 is a member) may be a significant source of UHECRs~\cite{Ding:2021emg}. 

\par 
Thanks to its multimessenger and multiwavelength complementarity, RNO-G represents an important piece of the multimessenger landscape over the coming decade. As will be discussed further in Section~\ref{sec:transient}, RNO-G will have the capability to follow-up multimessenger alerts from other observatories around the world. Due to the large number of excellent observing locations in the Northern hemisphere, these include ground-based observatories like HAWC~\cite{Mostafa:2013lth}, CTA~\cite{CTAConsortium:2017dvg}, and LHAASO~\cite{LHAASO:2019qtb} in gamma-rays and ZTF~\cite{2019PASP..131a8002B}, which has detected a number of potential tidal disruption events in optical wavelengths. RNO-G additionally shares a field-of-view with the Telescope Array, which primarily observes UHECRs, and IceCube at lower energies, as previously mentioned.

\begin{figure}
    \centering
    \includegraphics[width=0.9\textwidth]{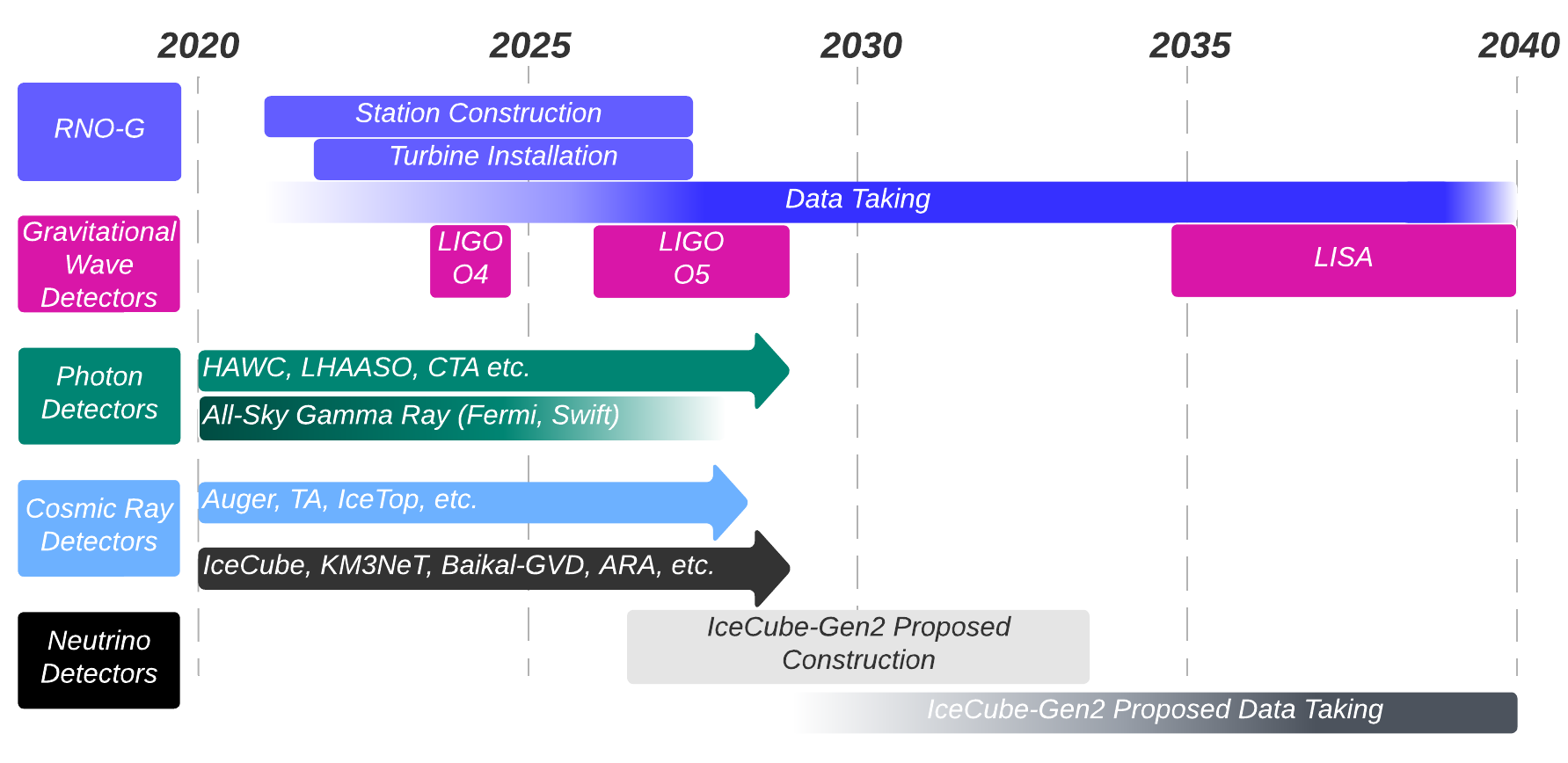}
    \caption{Expected timeline of multimessenger facilities over the next 15 years.}
    \label{fig:timeline}
    \vspace{-0.1in}
\end{figure}

\par
In particular, RNO-G's completion over the next few years places it at a particularly unique time for multimessenger science, as highlighted in Fig.~\ref{fig:timeline}. Firstly, the LIGO-Virgo-KAGRA (LVK) O4 run is currently underway and will continue until 2025, followed shortly thereafter by their O5 run from 2027 to 2030~\cite{ligoIGWNObserving}. At the same time, all-sky gamma-ray telescopes, such as \textit{Fermi} and Swift, are still operating but may be decomissioned in the early 2030s, leaving a GeV-gap gamma-ray sky monitoring~\cite{Engel:2022yig}. During this golden era for multimessenger science, RNO-G will be the only facility able to provide UHE neutrino follow-up to alerts. 

\section{Transient follow-up}\label{sec:transient}

\par 
In order to fully participate in the global multimessenger network, RNO-G will respond to real-time multimessenger alerts and, eventually, will issue its own alerts. RNO-G will monitor multimessenger alert networks, such as AMON~\cite{amon}, and respond to high-signalness alerts in one of two modes. 

\par
For most alerts, RNO-G will respond in \textit{normal mode}. In normal mode, each RNO-G station triggers at a rate of ${\sim}1$~Hz, limited by the wireless LTE throughput. In this mode, ``listeners'' subscribed to multimessenger alert streams will monitor for alerts with a high probability of being a UHE neutrino source, based on the alert's source type, distance, and other alert-specific parameters (e.g. the probability of an event being a binary neutron start merger $p_\mathrm{BNS}$ in LVK alerts). Once a high-neutrino-probability alert is identified an automated analysis is performed to search for neutrino events within an alert-specific time window $\Delta t$ around the alert time $t_0$. For events within this time window with a high neutrino probability, an initial reconstruction of the event direction and neutrino energy will be made. The timing, spatial, and energy information will be combined with the spatial and temporal information from the original alert to determine the signalness of the candidate neutrino and the probability of chance coincidence (i.e.\ false alarm rate). For candidates with sufficiently high signalness and low false alarm rate, a follow-up to the alert will be issued. 

\begin{figure}
    \centering
    \begin{subfigure}{0.49\textwidth}
        \includegraphics[height=1.8in]{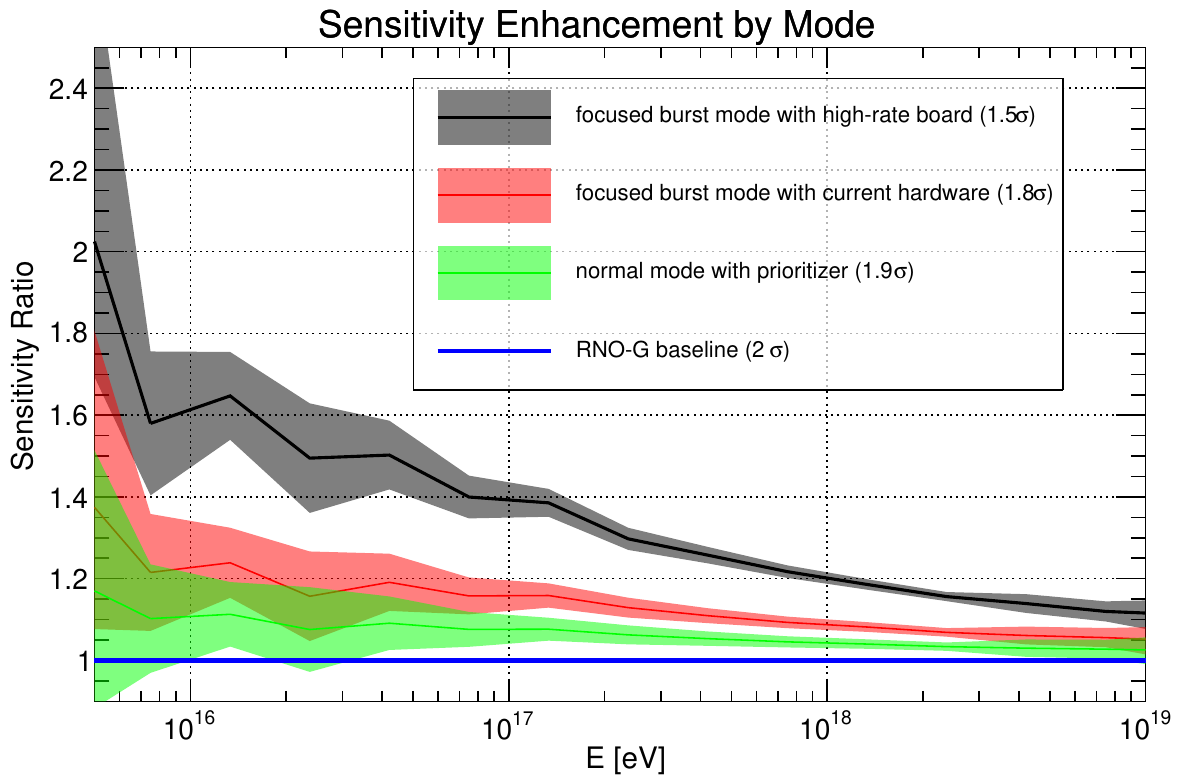}
        \caption{}\label{fig:improvement}
    \end{subfigure}
    \begin{subfigure}{0.49\textwidth}
        \includegraphics[height=1.8in]{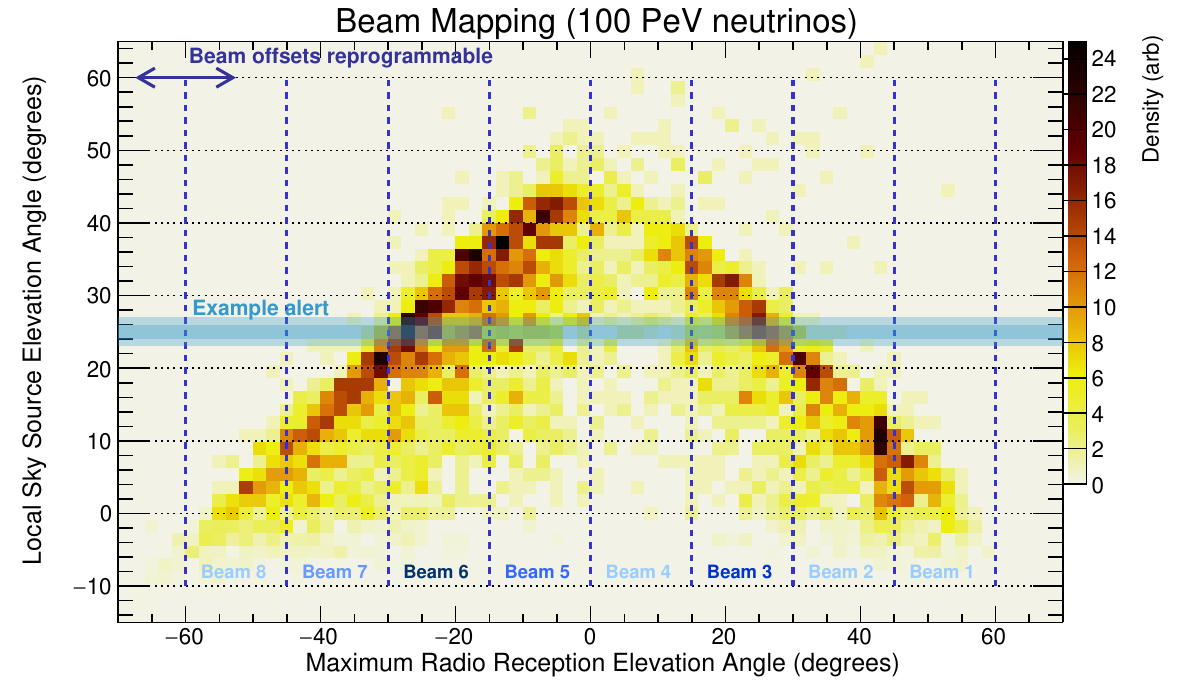}
        \caption{}\label{fig:beampointing}
    \end{subfigure}
    \caption{Left: Increase in instantaneous aperture in burst mode relative to baseline. Right: The mapping between radio reception direction and neutrino source direction. RNO-G's default beams are shown along with an example alert direction.}
    \label{fig:burstmode}
    \vspace{-0.2in}
\end{figure}

\par
For exceptional alerts, RNO-G will respond in \textit{burst mode}. Burst mode allows RNO-G to respond to high quality alerts in its current field-of-view with increased sensitivity. This is done by optimizing the trigger threshold in each of its beams for the alert to achieve the maximum possible trigger rate (up to ${\sim}100$ Hz) for signals correlated with the alert direction (see Fig.~\ref{fig:beampointing}). This temporarily boosts the instantaneous aperture (i.e.\ transient source sensitivity) up to twofold, with the largest increase at low energies (see Fig.~\ref{fig:improvement}). This is ideal since the neutrino flux is expected to be largest in the low energy end of RNO-G's sensitive range for most models of a transient neutrino flux~\cite{RNO-G:2020rmc}. After the burst period has concluded the same automated analysis described above will commence.

\section{Science with diffuse neutrinos}\label{sec:diffuse}

\begin{figure}
    \centering
    \includegraphics[width=\textwidth]{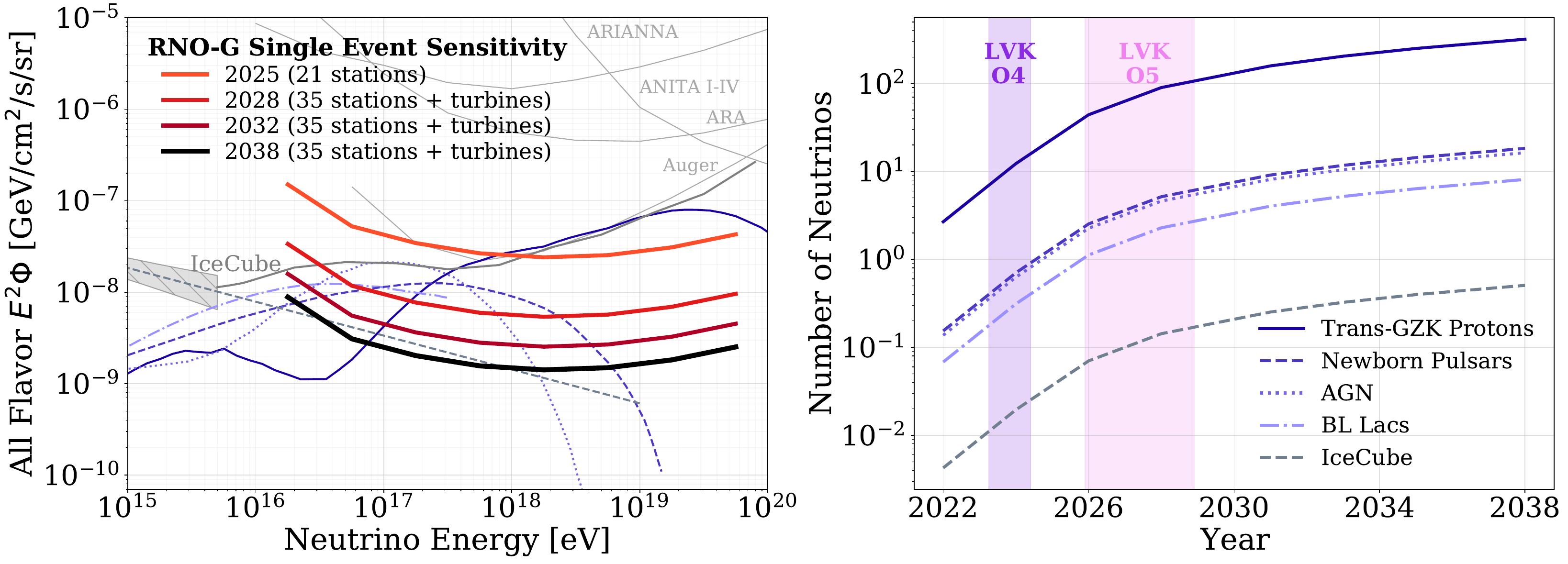}
    \caption{Left: Evolution of RNO-G's diffuse flux sensitivity over the next 15 years, along with several predictions for the neutrino flux from UHECR sources. Right: The expected number of observed neutrinos in the RNO-G livetime under different diffuse neutrino flux models~\cite{Muzio:2023skc,FangPulsar,Rodrigues:2020pli,Padovani:2015mba}.}
    \label{fig:diffuse_sensitivity}
    \vspace{-0.1in}
\end{figure}

\par 
Today, with only 7 stations currently operating, RNO-G is already the largest in-ice neutrino detector in the world by effective volume. Over the next decade it will become the world's most sensitive detector of UHE neutrinos. This places RNO-G in a prime position to make a strong impact on the study of both astrophysical neutrinos and UHECRs. RNO-G will either discover UHE neutrinos or place strong constraints on models of their sources. Figure~\ref{fig:diffuse_sensitivity} summarizes the evolution of RNO-G sensitivity to a diffuse neutrino flux over the next 15 years.

\par
Under the most optimistic models for the UHE neutrino flux, RNO-G could realistically discover a neutrino by 2025~\cite{vanVliet:2019nse,arianna200,Muzio:2023skc,Ehlert:2023btz}. If such a detection is made, RNO-G will immediately have discovered that there are trans-GZK CR sources at high redshifts (beyond $z{\sim} 1$) and that there is a significant flux of protons arriving at Earth above $30$~EeV~\cite{Muzio:2023skc}. Conversely, the lack of a neutrino detection by 2025 will place strong constraints on UHECR sources beyond the GZK horizon.

\par
By 2028, RNO-G's full detector array will have been deployed and its accumulated exposure will have significantly increased. If a neutrino is detected in its livetime through 2028 it would be a strong indication that high redshift UHECR sources are more luminous than those at low redshifts (i.e.\ that the UHECR source evolution is strongly positive). In particular, this would be strong evidence against negatively evolving sources, like BL Lacs, as being the sources of UHECRs~\cite{Rodrigues:2020pli}. However, if no neutrinos are detected by 2028 then RNO-G will place strong constraints on UHECR source models which assume a strong positive evolution or produce a significant proton fraction above $30$~EeV~\cite{Muzio:2023skc}.

\par
RNO-G's exposure by 2032 will be sufficient to probe the connection between the IceCube astrophysical neutrino flux and UHEs. Detection of a neutrino at this level of exposure would indicate a hardening of the astrophysical neutrino spectrum at UHEs. This would be a strong indication that sources of astrophysical neutrinos above and below $30$~PeV are largely divorced. Such a situation would imply that UHE neutrino detectors are required to fully understand the sources of neutrinos, since lower energy detectors will not have access to such a UHE source population. Conversely, a lack of neutrino detections would allow for the possibility that the sources of neutrinos seen by IceCube are also the most luminous sources of neutrinos at UHEs.

\par
Finally, by 2038 RNO-G will have enough data to start probing specific astrophysical UHECR source models. Neutrino detections, or lack thereof, with this level of exposure will give the first concrete evidence for or against AGN~\cite{Padovani:2015mba}, BL Lacs~\cite{Rodrigues:2020pli}, and newborn pulsars~\cite{FangPulsar} as the sources of UHECRs. Detection of neutrinos below ${\sim} 1$~EeV would provide strong evidence that UHECRs have a significant number of interactions inside their source environment -- a key uncertainty in UHECR source modeling. On the other hand, a lack of neutrinos would place strong constraints on the UHECR source luminosity-density, a recovery in the UHECR spectrum beyond $10^{20.3}$~eV, and the cutoff energy of the astrophysical neutrino spectrum. In addition to these specific constraints, RNO-G will have achieved world-record sensitivity to the diffuse neutrino flux from $10$~PeV-$100$~EeV.

\section{Probes of particle physics}\label{sec:particle}

\par
Observation of a UHE neutrino by RNO-G will represent the most energetic neutrino ever observed, far beyond the energies accessible by terrestrial facilities. As can be seen in Fig.~\ref{fig:diffuse_sensitivity}, in the most optimistic scenarios RNO-G could observe hundreds of neutrinos by 2038. This would enable RNO-G to become a leading probe of particle physics at the highest energies and beyond the Standard Model (BSM). 

\par
Observations of neutrinos will enable RNO-G to measure the neutrino-nucleon cross section at $\sqrt{s} \gtrsim 10$~TeV. This will allow RNO-G to probe BSM scenarios including extra dimensions, leptoquarks, and microscopic black hole production~\cite{Esteban:2022uuw,Valera:2022ylt,Ackermann:2022rqc}. Neutrino observations with RNO-G will also probe secret neutrino interactions, neutrino-dark matter (DM) interactions, and Lorentz invariance violation~\cite{Stecker:2014xja,Tomar:2015fha,Kelly:2018tyg,Ellis:2018ogq,Esteban:2021tub,Berryman:2022hds}. Even in the most pessimistic scenario of no neutrino observations by 2038, RNO-G will set strong limits on annihilating DM models~\cite{Guepin:2021ljb}.

\section{Conclusion}\label{sec:conclusion}

\par
The Radio Neutrino Observatory in Greenland (RNO-G) is poised to become the most sensitive observatory of ultrahigh energy (UHE) neutrinos in the world. With 7 of its 35 total stations currently deployed, it is already the largest in-ice neutrino detector ever constructed by effective volume. By combining the improved event reconstruction provided by surface antennas with the increased diffuse flux sensitivity provided by deep antennas, RNO-G will become a critical component of the multimessenger landscape over the coming decade. 

\par
RNO-G is the only UHE neutrino observatory in the world with a view of the Northern sky, home to a number of promising sources including NGC 1068. With its precise pointing and capability to temporarily boost its instantaneous aperture in the direction of transient events, RNO-G will provide critical information to the global multimessenger alert network. It will fulfill this role during a golden era of multimessenger science, when both the LIGO-Virgo-KAGRA gravitational wave network undergoes its run O4 and O5 and all-sky telescopes (such as \textit{Fermi} and Swift) continue to monitor the gamma-ray sky.

\par
RNO-G will provide critical probes of the astrophysical sources of UHECRs and their properties beyond the GZK horizon. The large effective area of RNO-G will allow for an unprecedented exposure to neutrinos in the $10$~PeV to $100$~EeV energy range. Detection or non-detection of neutrinos in this range will provide critical information on the evolution of UHECR sources, the proton fraction above $30$~EeV, and the significance of in-source UHECR interactions. Moreover, RNO-G will probe models of BL Lacs, AGN, and newborn pulsars as UHECR sources. Finally, the connection between the astrophysical neutrino flux and UHE neutrinos will be illuminated. 

\par
If Nature realizes an optimistic UHE neutrino flux, RNO-G will observe hundreds of neutrinos over the next 15 years. This opens the possibility for RNO-G to push the boundaries of particle physics, including the measurement of the neutrino-nucleon cross-section at unprecedented energies, and physics beyond the Standard Model. Even in the most pessimistic scenario, RNO-G will probe unexplored parts of the annihilating dark matter parameter space.

\begingroup
\setstretch{0.8}
\setlength{\bibsep}{2.5pt}
\bibliographystyle{ICRC}
\bibliography{references}
\endgroup

%

\clearpage

\section*{Full Author List: RNO-G Collaboration}
\scriptsize
\noindent
J. A. Aguilar$^{1}$, 
P.~Allison$^{2}$, 
D.~Besson$^{3}$, 
A.~Bishop$^{10}$, 
O.~Botner$^{4}$, 
S.~Bouma$^{5}$, 
S.~Buitink$^{6}$, 
W.~Castiglioni$^{8}$, 
M.~Cataldo$^{5}$, 
B.~A.~Clark$^{7}$, 
A.~Coleman$^{4}$, 
K.~Couberly$^{3}$, 
P.~Dasgupta$^{1}$, 
S.~de Kockere$^{9}$, 
K.~D.~de Vries$^{9}$, 
C.~Deaconu$^{8}$, 
M.~A.~DuVernois$^{10}$, 
A.~Eimer$^{5}$, 
C.~Glaser$^{4}$, 
T.~Gl{\"u}senkamp$^{4}$, 
A.~Hallgren$^{4}$, 
S.~Hallmann$^{11}$, 
J.~C.~Hanson$^{12}$, 
B.~Hendricks$^{14}$, 
J.~Henrichs$^{11, 5}$, 
N.~Heyer$^{4}$, 
C.~Hornhuber$^{3}$, 
K.~Hughes$^{8}$, 
T.~Karg$^{11}$, 
A.~Karle$^{10}$, 
J.~L.~Kelley$^{10}$, 
M.~Korntheuer$^{1}$, 
M.~Kowalski$^{11, 15}$, 
I.~Kravchenko$^{16}$, 
R.~Krebs$^{14}$, 
R.~Lahmann$^{5}$, 
P.~Lehmann$^{5}$, 
U.~Latif$^{9}$, 
P.~Laub$^{5}$, 
C.-H. Liu$^{16}$, 
J.~Mammo$^{16}$, 
M.~J.~Marsee$^{17}$, 
Z.~S.~Meyers$^{11, 5}$, 
M.~Mikhailova$^{3}$, 
K.~Michaels$^{8}$, 
K.~Mulrey$^{13}$, 
M.~Muzio$^{14}$, 
A.~Nelles$^{11, 5}$, 
A.~Novikov$^{19}$, 
A.~Nozdrina$^{3}$, 
E.~Oberla$^{8}$, 
B.~Oeyen$^{18}$, 
I.~Plaisier$^{5, 11}$, 
N.~Punsuebsay$^{19}$, 
L.~Pyras$^{11, 5}$, 
D.~Ryckbosch$^{18}$, 
F.~Schl{\"u}ter$^{1}$, 
O.~Scholten$^{9, 20}$, 
D.~Seckel$^{19}$, 
M.~F.~H.~Seikh$^{3}$, 
D.~Smith$^{8}$, 
J.~Stoffels$^{9}$, 
D.~Southall$^{8}$, 
K.~Terveer$^{5}$, 
S. Toscano$^{1}$, 
D.~Tosi$^{10}$, 
D.~J.~Van Den Broeck$^{9, 6}$, 
N.~van Eijndhoven$^{9}$, 
A.~G.~Vieregg$^{8}$, 
J.~Z.~Vischer$^{5}$, 
C.~Welling$^{8}$, 
D.~R.~Williams$^{17}$, 
S.~Wissel$^{14}$, 
R.~Young$^{3}$, 
A.~Zink$^{5}$
\\ 
\\ 
\noindent
$^1$ Universit\'e Libre de Bruxelles, Science Faculty CP230, B-1050 Brussels, Belgium \\ 
$^2$ Dept.\ of Physics, Center for Cosmology and AstroParticle Physics, Ohio State University, Columbus, OH 43210, USA \\ 
$^3$ University of Kansas, Dept.\ of Physics and Astronomy, Lawrence, KS 66045, USA \\ 
$^4$ Uppsala University, Dept.\ of Physics and Astronomy, Uppsala, SE-752 37, Sweden \\ 
$^5$ Erlangen Center for Astroparticle Physics (ECAP), Friedrich-Alexander-Universit\"{a}t Erlangen-N\"urnberg, 91058 Erlangen, Germany \\ 
$^6$ Vrije Universiteit Brussel, Astrophysical Institute, Pleinlaan 2, 1050 Brussels, Belgium \\ 
$^7$ Department of Physics, University of Maryland, College Park, MD 20742, USA \\ 
$^8$ Dept.\ of Physics, Enrico Fermi Inst., Kavli Inst.\ for Cosmological Physics, University of Chicago, Chicago, IL 60637, USA \\ 
$^9$ Vrije Universiteit Brussel, Dienst ELEM, B-1050 Brussels, Belgium \\ 
$^{10}$ Wisconsin IceCube Particle Astrophysics Center (WIPAC) and Dept.\ of Physics, University of Wisconsin-Madison, Madison, WI 53703,  USA \\ 
$^{11}$ Deutsches Elektronen-Synchrotron DESY, Platanenallee 6, 15738 Zeuthen, Germany \\ 
$^{12}$ Whittier College, Whittier, CA 90602, USA \\ 
$^{13}$ Dept.\ of Astrophysics/IMAPP, Radboud University, PO Box 9010, 6500 GL, The Netherlands \\ 
$^{14}$ Dept.\ of Physics, Dept.\ of Astronomy \& Astrophysics, Penn State University, University Park, PA 16801, USA \\ 
$^{15}$ Institut für Physik, Humboldt-Universit\"at zu Berlin, 12489 Berlin, Germany \\ 
$^{16}$ Dept.\ of Physics and Astronomy, Univ.\ of Nebraska-Lincoln, NE, 68588, USA \\ 
$^{17}$ Dept.\ of Physics and Astronomy, University of Alabama, Tuscaloosa, AL 35487, USA \\ 
$^{18}$ Ghent University, Dept.\ of Physics and Astronomy, B-9000 Gent, Belgium \\ 
$^{19}$ Dept.\ of Physics and Astronomy, University of Delaware, Newark, DE 19716, USA \\ 
$^{20}$ Kapteyn Institute, University of Groningen, Groningen, The Netherlands \\ 

\subsection*{Acknowledgments}

\noindent
We are thankful to the staff at Summit Station for supporting our deployment work in every way possible. We also acknowledge our colleagues from the British Antarctic Survey for embarking on the journey of building and operating the BigRAID drill for our project.
We would like to acknowledge our home institutions and funding agencies for supporting the RNO-G work; in particular the Belgian Funds for Scientific Research (FRS-FNRS and FWO) and the FWO programme for International Research Infrastructure (IRI), the National Science Foundation (NSF Award IDs 2118315, 2112352, 211232, 2111410) and the IceCube EPSCoR Initiative (Award ID 2019597), the German research foundation (DFG, Grant NE 2031/2-1), the Helmholtz Association (Initiative and Networking Fund, W2/W3 Program), the University of Chicago Research Computing Center, and the European Research Council under the European Unions Horizon 2020 research and innovation programme (grant agreement No 805486). M.S. Muzio thanks the NSF for support through the MPS-ASCEND Postdoctoral Fellowship under Award 2138121.

\end{document}